%% file: main.tex
\documentclass[prl,twocolumn,english,superscriptaddress,floatfix,longbibliography]{revtex4-2}

\usepackage{graphicx}
\usepackage{dcolumn}
\usepackage{bm}
\usepackage{amssymb}   
\usepackage{amsmath}
\usepackage{verbatim}
\usepackage{multirow}
\usepackage{xcolor} 
\usepackage{amsmath}
\usepackage{amsfonts}
\usepackage{soul}
\usepackage[pdfencoding=auto,psdextra]{hyperref}
\usepackage{makecell}
\hypersetup{colorlinks=true, pdfstartview=FitV, linkcolor=blue, citecolor=black, plainpages=false, pdfpagelabels=true, urlcolor=blue}
\usepackage[all]{hypcap}

\begin{document}

\title{Correlated quasiparticle poisoning from phonon-only events in superconducting qubits}

\author{E. Yelton}
\affiliation{Department of Physics, Syracuse University, Syracuse, NY 13244-1130, USA} 
\author{C. P. Larson}
\affiliation{Department of Physics, Syracuse University, Syracuse, NY 13244-1130, USA} 
\author{K. Dodge}
\affiliation{Intelligence Community Postdoctoral Research Fellowship Program, Department of Physics, Syracuse University, Syracuse, NY 13244-1130, USA}
\author{K. Okubo}
\affiliation{Department of Physics, Syracuse University, Syracuse, NY 13244-1130, USA} 
\author{B. L. T. Plourde}
\affiliation{Department of Physics, Syracuse University, Syracuse, NY 13244-1130, USA} 
\affiliation{Department of Physics, University of Wisconsin-Madison, Madison, Wisconsin 53706, USA}

\date{\today}


\begin{abstract}
Throughout multiple cooldowns we observe a power-law reduction in time for the rate of multi-qubit correlated 
poisoning events, while the rate of shifts in qubit offset-charge remains constant; evidence of a non-ionizing source of pair-breaking phonon bursts for superconducting qubits. We investigate different types of sample packaging, some of which are sensitive to mechanical impacts from the cryocooler pulse tube. One possible source of these events comes from relaxation of thermally-induced stresses from differential thermal contraction between the device layer and substrate.
\end{abstract}

%
%

\maketitle


Correlated errors in superconducting qubit arrays present a challenge to the implementation of quantum error correction. A significant source of these errors arises from background radiation, such as cosmic-ray muons and $\gamma$ rays~\cite{McEwen2021,Vepsalainen2020,Wilen2021,Cardani2021,Harrington2024,li2024}.
These ionizing impacts create a burst of electron-hole 
pairs in the Si substrate, which emit phonons when the electrons undergo intervalley scattering or the charge carriers recombine~\cite{Kelsey2023}. The energetic phonons spread throughout the substrate and generate excess quasiparticles (QPs) in the qubit electrodes. 
Additionally, the electrons and holes can be trapped via charge defects within the crystal substrate, changing the local charge environment~\cite{Wilen2021}.
Thus, QP poisoning from background radiation impacting superconducting qubit chips has both a characteristic charge response and pair-breaking phonon burst.
    

The development of radiation mitigation strategies is crucial for implementing quantum error correction.
One strategy involves reducing the overall flux of radiation incident on the device through shielding and choice of materials ~\cite{Cardani2021, Vepsalainen2020}. 
Other approaches rely on mitigation 
at the device level, such as phonon downconverting films, either with low-gap superconducting structures~\cite{Henriques2019,Karatsu2019} or normal-metal islands~\cite{Iaia2022,Yelton2024}, as well as superconductor gap engineering~\cite{Kamenov2023,Harrington2024,McEwan2024}. These on-chip techniques aim to reduce the effects of pair-breaking phonons on the tunneling of QPs that are generated in the energy cascade of a particle impact~\cite{Martinis2021}. 
Additionally, modified error correction protocols that are robust against these correlated events have been explored~\cite{Xu2022,Sane2023,Suzuki2022}. 
Besides ionizing radiation in the qubit environment, it is important to explore other potential sources of correlated QP poisoning. 
   
    
\begin{figure}[t!]
\centering
\includegraphics[width=3.4in]{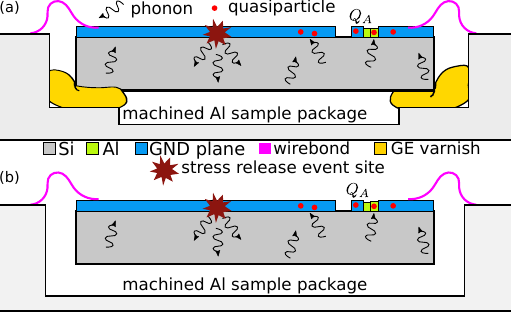}
\caption{
{\bf Diagram of stress release events.} Differential thermal contraction of the Si substrate, device films, and sample packaging creates stress fields within our device that can relax throughout the cooldown. We compare two chip mounting designs: (a) standard method, with the chip resting on a small ledge of the machined Al sample packaging and anchored to the box with GE varnish adhesive, and (b) suspended method, where the chip is suspended by an array of wirebonds. 
\label{fig:microfractures}}
\end{figure}
    
An experiment monitoring QP tunneling off of a mesoscopic superconducting island observed a decrease in tunneling rates as an inverse-time power law over the duration of the device cooldown ~\cite{Mannila2022}. Such a time-dependence is incompatible with poisoning from ionizing radiation.
A transition edge sensor (TES)-based 
experiment measured two devices: 
one attached with adhesive in the sample package, the other suspended from its mount by wirebonds. 
The experimenters observed a reduction in the rate of excess low-energy events 
for a device suspended by wirebonds compared to the device that was glued into the sample package with GE varnish; the rate of these low energy events decreased 
in time following the start of the cooldown, while the rates of higher energy events did not change for either device~\cite{Anthony2022}.
Based on these measurements, one proposed source of non-ionizing low-energy excess events is mechanical stress caused by differential thermal contraction of the device substrate, films, and sample packaging. The time dependence could be attributed to the relaxation of these stresses over the course of the cooldown. 
One suggested microscopic model for these events is the tunneling and subsequent transport of crystal dislocations in metal thin films~\cite{Romani2024}. 
Another proposed model involves localized disordered zones of the device substrate that release energy as they reconfigure to metastable states in an avalanche-like process~\cite{Nordlund2024}.
These low-energy excess events constitute the dominant background for many ongoing dark matter searches~\cite{Edelweiss2016, Hehn2016, Abdelhameed2019, SuperCDMS2021, Alkhatib2021, EXCESS2022, Anthony2022, CRESST-III2023}. 
Given how similar these detectors are to superconducting qubits, it is possible that such non-ionizing events could be present in both technologies. 

Both superconducting qubits and sensors for rare event searches 
use a single crystal substrate, typically Si, with superconducting thin-film elements on the 
surface of the substrate, operating at temperatures on the order of 10~mK. Generally, metals thermally contract much more than Si from room temperature to 4~K 
~\cite{Ekin}. The thermal stress at the substrate-film interface is the stress required to overcome the strain created by the differential thermal contraction between the device layer and substrate. For both Al and Nb films on Si, the thermal stress from room temperature to 4~K is on the order of hundreds of MPa (supplement \ref{sec:diff_thermal_contract}) \cite{Ekin, Namm1964, Pal-Val2006}.

In this manuscript, we present evidence of phonon bursts generated without a corresponding charge signal for several devices, each consisting of six-qubit arrays of charge-sensitive transmons on a Si substrate with both Nb and Al ground plane materials. 
We also explore the dependence of phonon-only events for two different sample packaging designs: (1) our conventional sample mount where the device rests on a machined ledge in an Al sample box and small amounts of GE varnish are added to the corners to anchor the device [Fig.~\ref{fig:microfractures}(a)]; (2) a suspended arrangement where the pocket has no direct contact with the device and the chip is held in place by an array of Al wirebonds [Fig.~\ref{fig:microfractures}(b)]. 
\begin{figure}[t!]
\centering
\includegraphics[width=\linewidth]{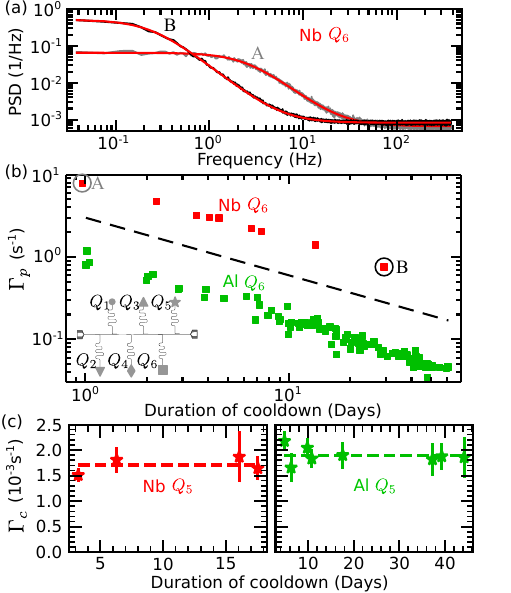}
\caption{
{\bf Evidence of non-ionizing events.} 
(a) Example power spectral densities for QP charge-parity measurements early (gray data, A) and late (black data, B) in the cooldown with corresponding fits (red).
(b) Characteristic QP parity switching rates $\Gamma_p$ 
for the same qubit, $Q_6$, for two devices with different ground plane materials: Nb (red) and Al (green). 
Dashed black line is a power law with exponent -0.7; labels A (B) refer to the early (late) data plotted in (a). (c) The large ($>0.15e$) offset-charge jump rates $\Gamma_c$ measured at different times in the cooldown for the same two devices in (b); left (right) corresponds to Nb (Al) ground plane device.
\label{fig:phonon_only}}
\end{figure}

Given the 
charge dispersion $\delta f$ ($\sim$2-12~MHz) for these qubits, we can utilize two measurement sequences to 
characterize the nature of QP poisoning
events in our qubits: QP charge-parity switching and charge tomography.
The transition frequency of charge-sensitive transmons depends on whether the total charge on the qubit island is an odd or even multiple of $e$. Changes in the QP charge-parity state can be attributed to a QP tunneling off or onto the island through the Josephson junction. Thus, an excess in QP density $n_{qp}$ correlates with a higher probability of a charge-parity switching event \cite{Catelani2011, Marchegiani2022}.
Using a standard QP charge-parity mapping protocol: $X/2$ pulse
, idle for $1/4\delta f$, 
$Y/2$ pulse, measure, we map the QP parity state to the qubit readout signal axis
\cite{Riste2013,Serniak2018,Christensen2019,Iaia2022}. 
Repeating this sequence on a fixed measurement interval $\Delta t$ and digitizing the parity signal via a simple thresholding technique, we extract a characteristic switching rate of the QP charge-parity state $\Gamma_p$ by fitting a Lorentzian to the power spectral density (PSD) of the digital parity signal [Fig.~\ref{fig:phonon_only}(a)], as done in Ref.~\cite{Riste2013}.

In previous work involving similar arrays of charge-sensitive transmons, we observed elevated charge-parity switching rates immediately after an 
unintentional warm-up of the cryostat. Upon cooling back down, we found that 
$\Gamma_p$ was 
elevated compared to measurements before the power outage \cite{Iaia2022}.
Here, we 
repeatedly 
measure QP charge-parity switching throughout multiple cooldowns and observe $\Gamma_p$ values that exhibit an inverse time power-law decrease [Fig.~\ref{fig:phonon_only}(b)].
This 
reduction in 
$\Gamma_p$ 
is consistent with the reduction of QP tunneling rates reported in Ref.~\cite{Mannila2022}.
In Fig.~\ref{fig:phonon_only}(b) we plot this power-law reduction in 
$\Gamma_p$ for qubits with a  
Nb (red) and Al (green) ground plane using our standard 
chip mounting.

To understand the overall lower $\Gamma_p$ for the device with the Al ground plane compared to the Nb chip, we consider the superconducting energy gap for the different elements on the Al chip. 
For Al films, the energy gap $\Delta_{\rm Al}$ depends inversely on film thickness~\cite{Marchegiani2022}. We fabricate our ground plane such that the energy gap of the ground plane is less than the energy gap in the qubit junction electrodes, which were fabricated in a Dolan bridge shadow-evaporation process~\cite{Dolan1977} (supplement \ref{sec:gap_engineering_Al}). 
This configuration provides some degree of phonon mitigation since phonons of pair-breaking energy relative to the energy gap of the qubit junction electrodes can downconvert in the ground plane~\cite{Henriques2019,Karatsu2019,Martinis2021}. Even with 
this phonon mitigation strategy, we still observe a power-law decrease in 
$\Gamma_p$ throughout the duration of the cooldown for both devices, indicating that a similar mechanism partially limits the QP charge-parity switching rates for devices with both Al and Nb ground planes. 


To determine if this reduction in 
$\Gamma_p$ could be attributed to ionizing radiation in the sample environment, we use our charge-sensitive transmons as electrometers of charge in the device substrate, as performed in Ref.~\cite{Christensen2019,Wilen2021,Larson2025} to detect impacts of high-energy particles on qubit devices. 
We adjust the dc voltage bias on a gate electrode near the qubit island to sweep the 
offset charge $n_g^{\rm ext}$ coupled to the qubit. 
For each bias point, we 
perform a charge tomography sequence: $X/2$ pulse, idle for $1/2\delta f$, $X/2$ pulse, measure. 
The resulting qubit one-state probability as a function of the total offset charge $n_g$ can be described by: $P_1= \frac{1}{2} \left[d+\nu\cos(\pi\cos(2\pi n_g))\right]$, where $d$ and $\nu$ are fit parameters to the signal axis, $n_g = n^{ext}_g+\delta n_{g}$, where 
$\delta n_g$ is the environmental offset charge~\cite{Christensen2019}.
We repeat this pulse sequence to monitor the rate $\Gamma_c$ of large ($>0.15e$) discrete shifts in $\delta n_g$ over time.
We repeat this experiment throughout the respective cooldowns for the Al and Nb ground plane devices. For both devices, the measured $\Gamma_c$ values remain constant, consistent with the average rate over the entire cooldown, $1.7(1)\times 10^{-3}\,$s$^{-1}$ ($1.9(1)\times 10^{-3}\,$s$^{-1}$) for the Nb (Al) ground plane devices [Fig.~\ref{fig:phonon_only}(c)]. 
The constant level of $\Gamma_c$ throughout the cooldown indicates a constant rate of ionizing impacts on both chips. Therefore, the elevated $\Gamma_p$ and subsequent decrease following the start of the cooldown must be due to non-ionizing events without a corresponding charge signal.

\begin{figure}[t!]
\centering
\includegraphics[width=\linewidth]{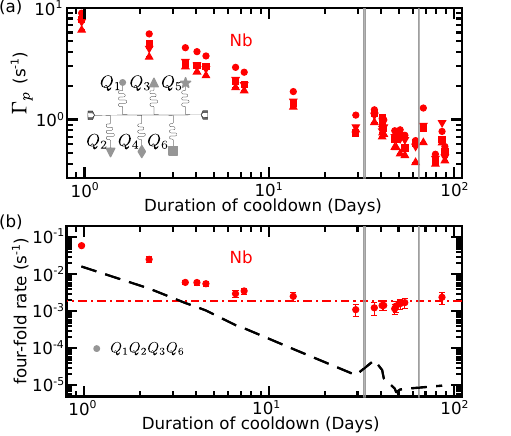}
\caption{
{\bf Correlated event rate during cooldown.} 
(a) 
$\Gamma_p$ for four qubits ($Q_1$,$Q_2$,$Q_3$,$Q_6$) on the Nb ground plane device throughout the cooldown. (b) 
4-fold correlated charge-parity 
switching rates of these four qubits throughout the cooldown. Dot-dashed red line is the predicted 4-fold saturation level from $R_{\rm impact}/2^4$. Dashed black line represents the expected random 4-fold coincident background computed from the single-qubit $\Gamma_p$ data from (a). Vertical lines represent thermal events in the cryostat, rising above, then back below, 100~mK. 
\label{fig:corr_parity}}
\end{figure}


In general, enhanced QP charge-parity switching rates correspond to excess QP density $n_{qp}$ in the qubit junction electrodes, which can be generated by both pair-breaking photons and phonons~\cite{Kaplan1976, Catelani2011, Wang2014, Houzet2019,deGraaf2020, Marchegiani2022, Diamond2022}. 
Previously, we have used the same cryostat and similar device design described in the present manuscript to 
study the behavior of 
back-side Cu islands, which reduced the $\Gamma_p$ levels by an order of magnitude compared to a device with no Cu islands \cite{Iaia2022}. This indicated that our experimental setup is likely phonon-limited, and thus is relatively insensitive to background infrared (IR) photons. 
In addition to best practices for IR shielding and filtering of signal leads~\cite{Connolly2024}, we attribute this insensitivity to IR radiation to the compact design of the qubit island, such that its spurious antenna resonance is a poor match to the free-space electromagnetic environment \cite{Liu2024}. 
More evidence of the dominance of phonon-mediated poisoning in Ref.~\cite{Iaia2022} came from measuring the rates of multi-qubit QP charge-parity switching, which were significantly reduced by the presence of the Cu islands. Without phonon mitigation, 
a burst of pair-breaking phonons within the device substrate, for example, from a high-energy particle impact, can efficiently spread throughout the entire substrate, poisoning multiple qubits simultaneously~\cite{Iaia2022, Yelton2024}. In contrast, 
correlated QP charge-parity switching is unlikely to be caused by 
photon-mediated poisoning due to the stochastic nature of photon absorption by qubits from the IR background \cite{Liu2024}. 

Using the same QP charge-parity mapping sequence for characterizing $\Gamma_p$, we monitor the QP charge-parity state on multiple qubits simultaneously. 
Repeating this measurement 
throughout the cooldown, we find the rate of coincident 
QP charge-parity switching is higher than the expected random coincident switching background (supplement~\ref{sec:corr_parity}). 
Additionally, we observe the rate of correlated multi-qubit QP charge-parity switching events also decreases with an inverse time power-law before saturating at long times at a level consistent with the environmental radiation background in our lab (see below). In Fig.~\ref{fig:corr_parity}(b) we plot the rate of 4-fold correlated QP charge-parity switching during the cooldown for the Nb ground plane device; 
the expected random coincident background rate is plotted as a black dashed line. In Fig.~\ref{fig:corr_parity}(a) we plot the 
$\Gamma_p$ values of the four qubits that comprise the 4-fold correlated event rates from Fig.~\ref{fig:corr_parity}(b). 
Since the rate of correlated QP charge-parity switching decreases with time while the rate of large offset-charge bursts in the substrate remains constant [Fig.~\ref{fig:phonon_only}(c)],
we conclude that the elevated QP charge-parity switching and subsequent decrease following the start of the cooldown are caused by phonon-only events that have no corresponding charge signal. 

Throughout the manuscript, we have defined Day 0 of the cooldown to be the point when the mixing chamber of the cryostat falls below 100~mK. During the cooldown we had an unexpected blockage in the cold trap of the dilution circuit, 
and as a result, the mixing chamber increased above 100~mK to $\sim$14~K. 
The temperature rise and subsequent recovery below 100~mK are both demarcated as vertical lines in Fig.~\ref{fig:corr_parity}, occurring $\sim$32 days after the start of the cooldown. There was a second warm-up above 100~mK $\sim$65 days into the cooldown from intentionally turning off the cryocooler pulse tube (details follow).
In Fig.~\ref{fig:corr_parity}(a) there is a clear increase in 
$\Gamma_p$ at these points. Nonetheless, these rates resume the decrease in time on an inverse time power law following the temperature spikes. However, the four-fold rates remain flat during this time after the temperature event on the mixing chamber [Fig.~\ref{fig:corr_parity}(b)]. The 4-fold rates saturate to 1.5(4)~$\times10^{-3}\,$s$^{-1}$ and are unaffected after the temperature rise and recovery in the cryostat. From the charge-sensing analysis in Ref.~\cite{Larson2025}, our qubits sense charge jumps $>0.15e$ due to particle impacts within $1$~mm from the qubit islands. For the $8 \times8~$mm$^2$ chip area and the average offset-charge jump rate $\Gamma_c$ from Fig.~\ref{fig:phonon_only}(c), we estimate the rate of ionizing particle impacts on the entire chip to be $R_{\rm impact} = 3(1)\,\times10^{-2}\,$s$^{-1}$. 
Since we can only detect half of the QP poisoning events on a given qubit (supplement~\ref{sec:corr_parity}), we expect the rate of 4-fold correlated QP charge-parity switching due to ionizing radiation to be $R_{\rm impact}/2^{4}\,$= 1.9(6)~$\times10^{-3}\,$s$^{-1}$ [Dot-dashed red line Fig.~\ref{fig:corr_parity}(b)].
Thus, after a sufficiently long cooldown, the rate of system-wide phonon-only events decreases below the level 
corresponding to background ionizing radiation. 

\begin{figure}[t!]
\centering
\includegraphics[width=\linewidth]{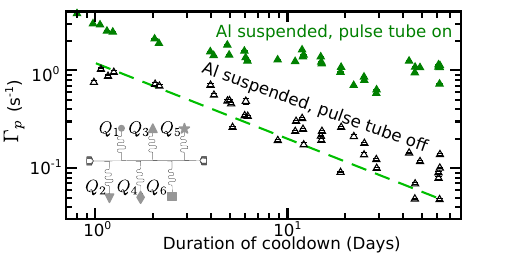}
\caption{
{\bf Suspended device with pulse tube on and off.}
$\Gamma_p$ for $Q_3$ of the Al ground plane device in the suspended sample packaging with the pulse tube on (dark green) and off (black open symbols). Power law fit of $\Gamma_p$ versus time for $Q_6$ in the conventional packaging from Fig.~\ref{fig:phonon_only}(b) is shown as a green dashed line.
\label{fig:pulse_tube}}
\end{figure}

In the absence of background radiation, these phonon-only events would remain a source of correlated errors in superconducting qubit arrays. Thus, it is important to understand the underlying physical mechanism 
so that devices can be designed to mitigate these events. 
To investigate this, we perform an  
experiment with a suspended qubit chip, similar to the TES measurements in Ref.~\cite{Anthony2022}. We use the same Al ground plane sample that we originally measured in our standard sample packaging (Fig.~\ref{fig:phonon_only}), but now mount the chip so that it is fully suspended by wirebonds around the edge of the chip [Fig.~\ref{fig:microfractures}(b)]. 

For the Al ground plane device cooled in the suspended configuration, we observe significantly elevated $\Gamma_p$ with a more modest reduction in time compared to the conventional sample packaging (Fig.~\ref{fig:pulse_tube}). 
To determine if this elevated rate is due to mechanically driven modes from the compression cycle of the cryocooler pulse tube, 
we measure 
$\Gamma_p$ 
with the pulse tube off for 7-minute intervals. 
During this time, the mixing chamber remains at its base temperature of $\sim$7~mK.
We then turn the pulse tube back on and allow the cryostat to recover for at least 1 hour (supplement~\ref{sec:pulse_tube_off}). 
Before the experiment, we 
measure $\Gamma_p$ 
to get a baseline 
before turning the pulse tube off. 
By day 49 of the cooldown, $\Gamma_p$ for the pulse tube on [off] is 1.07(3) [0.108(2)] s$^{-1}$. We observe this reduction in 
$\Gamma_p$ depending on the state of the pulse tube 
on all qubits in the array, but do not observe this dependence for a qubit chip in the conventional sample packaging (supplement~\ref{sec:pulse_tube_off}). This indicates that mechanical perturbations generated by the pulse tube operation and subsequent deformation of the suspending wirebonds, consistent with finite element modeling of wirebond deflections~\cite{Logg2012, Alnaes2015}, are likely the source of pair-breaking phonons that limit 
$\Gamma_p$
(supplement~\ref{sec:mechanical_modes}). In Ref.~\cite{Kono2024}, the authors found that the pulse tube operation of their cryostat drove spurious qubit state excitations; we have not observed any correlation in our qubit one-state occupation with the operation of our pulse tube. For the measurements in Ref.~\cite{Kono2024}, the device was mounted with GE varnish, thus their observations may involve a different physical mechanism.

We repeat this experiment with the pulse tube turned off for $\sim$7-minute intervals throughout the cooldown. We find that $\Gamma_p$ for the pulse tube off data 
follows a similar inverse time power law that was observed when the same device was mounted with GE varnish (Fig.~\ref{fig:pulse_tube}). 
This is evidence that the source of the phonon-only events we observe in our devices is likely dominated by thermal stress-relaxation events at the film-substrate interface, rather than the substrate-GE varnish interface for our devices in the conventional packaging. 
%
%
For our conventional sample packaging, we only apply small spots of GE varnish at the chip corners; by contrast, in Ref.~\cite{Anthony2022} it appears the adhesive was applied to the entire back side of the chip. Thus, it could be 
the stress field experienced by the chip is different between our measurements and those in 
Ref.~\cite{Anthony2022}.
These measurements highlight the importance of sample packaging for understanding the prevalence of phonon-only events in both superconducting qubit and dark matter detection devices.

This work presents evidence that these events are non-ionizing and mediated by phonon bursts within the substrate, consistent with the excess low energy events seen in rare event detectors~\cite{Hehn2016,Abdelhameed2019,Alkhatib2021,EXCESS2022}.
The inverse time power-law reduction in QP poisoning rates that we and others~\cite{Mannila2022} observe is reminiscent of stick-slip events, for example, from microfractures at the metal film-substrate interface or in the Si chip~\cite{Anthony2022, Romani2024, Nordlund2024}. 
We observe phonon-only events for devices with both Nb and Al ground plane materials. Given that, in general, metals thermally contract more than Si ~\cite{Ekin}, most superconducting qubit and detector arrays will be susceptible to this mechanism of phonon bursts, posing a source of correlated qubit errors and detector background events. Thus, characterizing and mitigating these bursts is crucial for realizing fault-tolerant quantum computers and extending the sensitivity of phonon-mediated sensors.

This work is supported by the U.S. Government under ARO grant W911NF-22-1-0257. Fabrication was performed in part at the Cornell NanoScale Facility, a member of the National Nanotechnology Coordinated Infrastructure (NNCI), which is supported by the National Science Foundation (Grant NNCI-2025233). We acknowledge useful discussions with R. McDermott
\input{bib_output_main.tex}
\widetext
\clearpage
\begin{center}
\textbf{\large Supplementary Information: Correlated quasiparticle poisoning from phonon-only events in superconducting qubits}
\end{center}
\def\thesection{\Roman{section}.}
\setcounter{secnumdepth}{3}
\setcounter{equation}{0}
\setcounter{figure}{0}
\setcounter{table}{0}
\setcounter{page}{1}
\renewcommand{\theHtable}{Supplement.\thetable}
\renewcommand{\theHfigure}{Supplement.\thefigure}
\makeatletter
\renewcommand{\theequation}{\arabic{equation}}
\renewcommand{\thefigure}{\arabic{figure}}
\renewcommand{\thetable}{\arabic{table}}
\newif\if@seccntdot
\def\@seccntformat#1{%
  \csname the#1\endcsname
  \if@seccntdot .\fi
  \quad
}
\renewcommand{\theequation}{S\arabic{equation}}
\renewcommand{\thefigure}{S\arabic{figure}}
\renewcommand{\thetable}{S\arabic{table}}
\renewcommand{\bibnumfmt}[1]{[S#1]}
\renewcommand{\citenumfont}[1]{S#1}

\input{RawText_supplement.tex}
\end{document}

%% file: RawText_supplement.tex
\renewcommand{\thesection}{\Roman{section}}
\section{Device Fabrication}
\label{sec:device_fab}

Both samples are fabricated on high-resistivity ($>$10 k$\mathrm{\Omega}$-cm) 100-mm Si wafers. Both ground planes are sputtered films of thickness 70-nm for the Nb device and 185-nm for the Al device. The Nb ground plane sample is patterned, developed, and dry-etched as in Ref.~\cite{sIaia2022, sLarson2025}, resulting in X-mon-style qubit islands with a 5-$\mu$m gap between the island and ground plane. The Al ground plane sample is patterned and developed similarly, but then undergoes a wet etch with Transene Aluminum Etchant type A (mixture of phosphoric, acetic, and nitric acid). Following etching, the Nb base-layer pattern includes alignment marks for performing aligned exposure for the electron beam lithography to define the Josephson junctions. However, because 
of the low electron microscopy contrast between Al and Si, 
the Al sample requires another lithography step to define Nb alignment marks on top of the patterned Al ground plane. We spin LOR3A lift-off resist and then DUV210-0.6 photoresist before exposing the alignment marks. After development, 100~nm of Nb was sputtered before being submerged in 1165 Remover (N-Methly-2-pyrrolidone) at 70$^{\circ}$ C for two 30-minute baths and cleaned in an ultrasonic bath to lift off the excess Nb and resist. With Nb alignment marks on both samples, we define Dolan-bridge Josephson junctions~\cite{sDolan1977} using electron beam lithography and then use electron beam evaporation of Al to create bottom (top) films of thickness 40~nm (80~nm), as in Ref.~\cite{sIaia2022}.

\section{Cooldown Details}
\label{sec:cooldown_details}

Our experiments span across three separate cooldowns with each cooldown focusing on a specific device. Table~\ref{tab:cooldowns} shows the details and timeline for each cooldown described in the main paper. The fridge configuration is identical to that in Ref.~\cite{sIaia2022, sLarson2025}, as indicated by the table.

\begin{table}[h]
    \centering
    \begin{tabular}{|c|c|c|c|}
        \hline
         Cooldown &  Duration (days) & Device & Configuration \\
         \hline
         1 & 61 & Al ground plane (conventional) & Ref.~\cite{sIaia2022} \\
         \hline
         2 & 103 & Nb ground plane (conventional) & Ref.\cite{sLarson2025} \\
         \hline
         3 & 110 & Al ground plane (suspended) & Ref.\cite{sLarson2025} \\
         \hline
    \end{tabular}
    \caption{\textbf{Cooldown Details.}}
    \label{tab:cooldowns}
\end{table}

\section{ Differential thermal contraction}
\label{sec:diff_thermal_contract}

As mentioned in the main text, metals thermally contract much more than Si from room temperature to 4~K;
for Al, the fractional change in length from room temperature to 4~K, $\Delta L/L$, is 41.5$\times10^{-4}$, for Nb, $\Delta L/L$ is 14.3$\times10^{-4}$,
while for Si, 
$\Delta L/L$ is 2.2$\times10^{-4}$~\cite{sEkin}.
The thermal stress 
is defined by the stress required to overcome the strain created by the thermal contraction between the ground plane and the device substrate, $\sigma = E(\Delta L/L),$ where $E$ is the Young's modulus of the material~\cite{sEkin}. Nb and Al have similar values of $E$ at cryogenic temperatures, $\sim$100~GPa~\cite{sNamm1964, sPal-Val2006}. For an Al (Nb) film on Si, the thermal stress from room temperature to 4~K is $\sim$ 450 MPa ($\sim$ 200 MPa)  \cite{sEkin, sNamm1964, sPal-Val2006}. Note that this estimate is only due to the differential thermal contraction between the films and the substrate and does not consider a contribution of inherent film stress from the deposition of the films. We leave the study of film stress from deposition and the resulting prevalence of these phonon-only events as a subject of future study. 

\section{ Details on QP charge-parity measurements}
\label{sec:parity}

The standard QP charge-parity mapping protocol is as follows: first we apply an $X/2$ pulse at the degeneracy frequency of the qubit $f_{01}$, idle for $1/4\delta f$, allowing the odd (even) state to precess $\pi/2$ (when at the maximum charge dispersion) about the equator of the Bloch sphere, then apply a final $Y/2$, which maps the odd (even) state to the qubit readout signal axis. \cite{sRiste2013,sSerniak2018,sChristensen2019,sIaia2022}. Repeating this experiment on a fixed measurement interval $\Delta t$, we can rapidly monitor the charge-parity state of the qubit island. 

Following Ref.~\cite{sRiste2013}, we digitize the QP charge-parity data with a simple thresholding scheme. The power spectral density (PSD) of the digital data can be characterized as random telegraph noise where the PSD can be described with a single Lorentzian, following Ref.~\cite{sRiste2013}:
\begin{equation}
S_{p}(f)=\frac{4F^2\Gamma_{p}}{(2\Gamma_{p})^2+(2\pi f)^2}+\left( 1-F^2\right)\Delta t,
\label{eq:Lorentzian_PSD}
\end{equation}
where $\Gamma_{ p}$ is the characteristic parity switching rate and $F$ is the parity sequence mapping fidelity. Note that this Lorentzian assumes that the digital signal switches between 1 and -1 \cite{sRiste2013}. 

\begin{figure}[h]
\centering
\includegraphics[width=\linewidth]{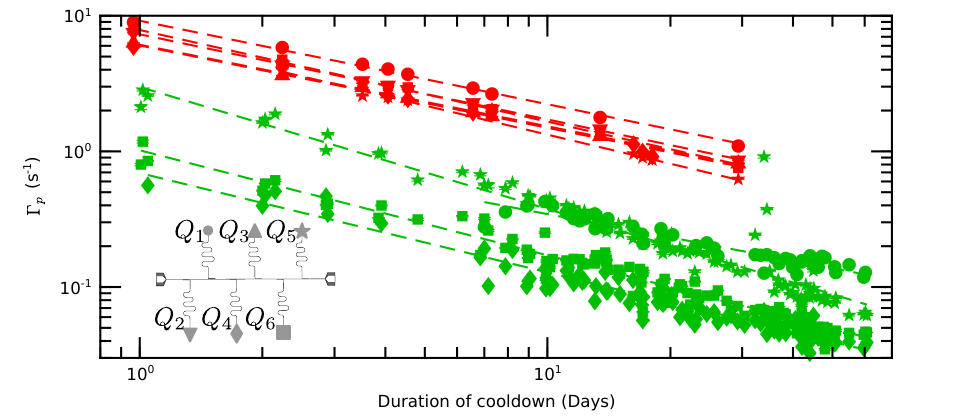}
\caption{
{\bf $\Gamma_p$ for the conventional samples.}
The characteristic QP charge-parity switching rates $\Gamma_p$ for the Nb (red) and Al (green) ground plane devices for multiple qubits on each device. Power law fits for the Nb (Al) ground plane devices are shown as red (green) dashed lines. 
\label{fig:PSD_fits}}
\end{figure}

In Fig.~\ref{fig:PSD_fits} we report the characteristic charge-parity switching rate $\Gamma_p$ for multiple qubits on both the Nb and Al ground plane devices that were mounted in the conventional sample packaging [a subset of qubits was shown in Fig.~\ref{fig:phonon_only}(b)]. We observe that the QP charge-parity switching rates versus time are best described with a power law $\Gamma_p(t) = At^{\alpha}$, where $t$ is the duration of time that the device has been at the base temperature. We find the Al [Nb] data are described with $\alpha = -0.7(1)~[-0.64(4)],$ similar to the time dependence of the QP tunneling rates reported in Ref.~\cite{sMannila2022}. The difference in these exponents could be due to the different material properties of the metal films or the device substrate \cite{sRomani2024, sNordlund2024} and is a subject of further study. 

We note that one of the qubits, $Q_5$, from the Al ground plane device [Fig.~\ref{fig:PSD_fits}(green stars)] experiences an increase and subsequent recovery in the QP charge-parity switching rate at $\sim30$ Days into the cooldown. We cannot attribute this rise in $\Gamma_p$ for this one qubit to a temperature rise of the cryostat since the cryostat stages remained stable during this time. Additionally, we would expect an increase in all of the qubits, as seen in Fig.~\ref{fig:corr_parity}, if the temperature rose in the cryostat. It is plausible that this increase in $\Gamma_p$ for this one qubit could be attributed to a quasiparticle-coupled two level system (TLS) \cite{sdeGraaf2020}. 
The physical mechanism of this increase in the parity switching rates has affected only this one qubit and is not characteristic of the phonon-only events described in the main text. As a result, we did not include these three data points in determining the power-law fit to the data. 

\section{ Gap engineering of the Al ground plane device}
\label{sec:gap_engineering_Al}

\begin{figure}[h]
\centering
\includegraphics[width=\linewidth]{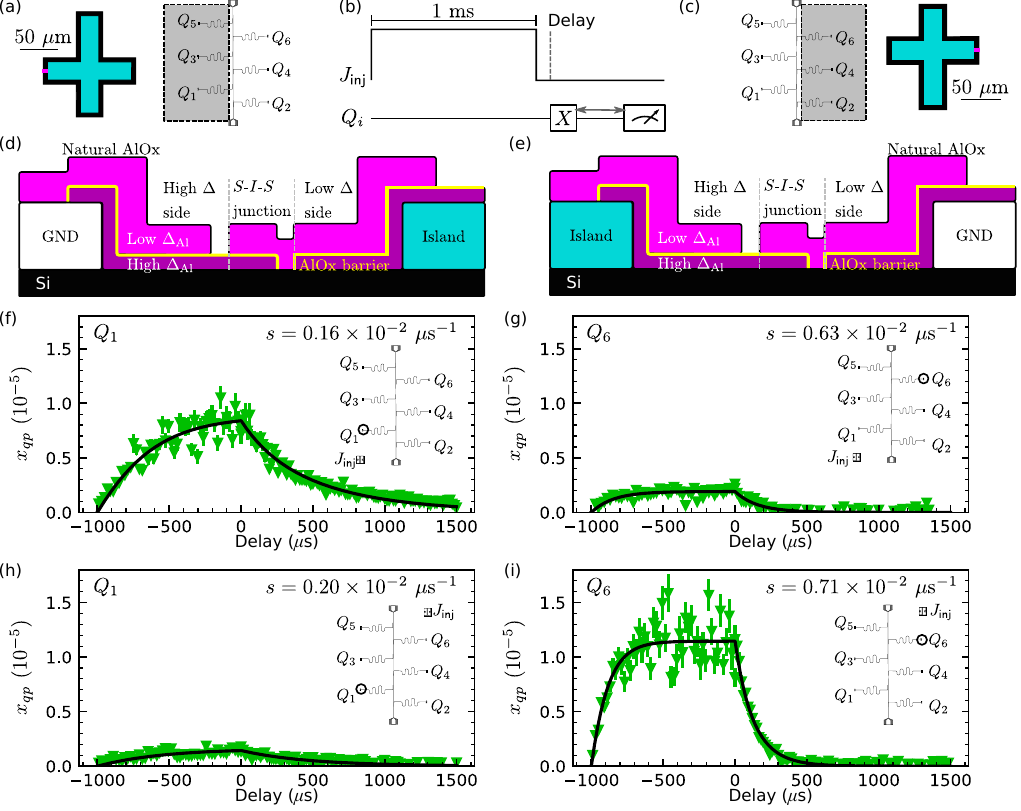}
\caption{
{\bf Phonon injection in the Al ground plane conventional sample.} (a) Diagram of qubit island geometry (cyan) for the non-ideal gap engineered junctions. The ground plane is the implied white space and the absence of a metal film is drawn in black. Throughout this figure the qubit layout is rotated 90\textdegree counterclockwise from the layout presented in other figures of the main manuscript. Gray shaded region indicates all non-ideal gap engineered qubits on this device ($Q_1,Q_3,Q_5$). (b) Phonon injection pulse sequence where we bias an on-chip tunnel junction $J_{\rm inj}$ for 1~ms and then wait for some delay before performing a $T_1$ measurement on $Q_i$. (c) Diagram of qubit island geometry for the gap engineered qubits, which are to the right of the common feedline ($Q_2,Q_4,Q_6$), highlighted by a gray shaded region. Color coding is identical to part (a). (d) Diagram of Josephson junction Al bilayer for the non-ideal gap-engineered qubits. Note that this drawing is not to scale. Black indicates the Si substrate, dark magenta is the thin, high-$\Delta$ Al film, and magenta indicates the thicker, low-$\Delta$ film. The junction's insulating oxide barrier is drawn in yellow. The high-$\Delta$ side of the junction (left side) connects to the ground plane, white, and the low-$\Delta$ side of the junction (right side) connects to the qubit island, cyan. (e) Diagram of gap-engineered qubits' junction bilayer with identical color coding as in part (d). The high-$\Delta$ side of the junction (left side) connects to the qubit island, cyan, and the low-$\Delta$ side of the junction (right side) connects to the ground plane, white. (f-i) Experimental results for phonon injection experiments indicated by green symbols; black curves represent a model of the QP density in the qubit junctions. Location of injector junction $J_{\rm inj}$ and the measured qubit (black circle) are shown in the insets of each figure. Biasing the injector junction located in the bottom left of the device and measuring the $x_{qp}$ response of (f) $Q_1$ (non-ideal gap engineered) and (g) $Q_6$ (gap engineered). Similarly, injection from the top right of the device and the resulting $x_{qp}$ response (h) $Q_1$ (non-ideal gap-engineered) and (i) $Q_6$ (gap-engineered). The trapping rates from the modeling are shown above the injector-qubit location insets in (f-i). These trapping rates are consistent regardless of injection location and can be attributed to the difference in the gap engineering of the qubits.}
\label{fig:Al_inj}
\end{figure}
    
For the Al ground plane data in Fig.~\ref{fig:PSD_fits}, there is a separation of the charge-parity switching rates where two of the four qubits are generally higher than the others. As mentioned in the main manuscript, the superconducting gap energy of Al $\Delta_{\textrm{Al}}$ depends on the thickness $d$ of the film, where $\Delta_{\textrm{Al}}(d)= \Delta_{bulk}+ad^{-1}$, where $\Delta_{bulk}$ is 180~$\mu$eV and $a$ is 600~$\mu$eV$\cdot$nm~\cite{sMarchegiani2022}. The Al ground plane has a thickness of $\sim$185~nm, which coincides with a superconducting gap $\Delta$ $\sim$183~$\mu$eV, which is lower than the superconducting gap of the Josephson junction electrodes of thickness 80~nm (40~nm) with corresponding gaps of $\sim$188~$\mu$eV ($\sim$195~$\mu$eV). 
        
The variation in the QP charge-parity switching rates amongst the Al ground plane qubits likely arises from the orientation of the asymmetric superconducting gap bilayer of the Josephson junction for each qubit with respect to the low-gap side of the junction being oriented toward (away from)
the qubit island. A similar effect was reported in Ref.~\cite{sHarrington2024} for a comparable qubit capacitor design consisting of an Al ground plane and an asymmetric junction bilayer.
        
Following the shadow evaporation process of the Dolan bridge depositions \cite{sDolan1977}, there is an ancillary low-gap film that sits on top of the high-gap side of the junction that can serve as a QP trap. As a result, there will be an abundance of QPs occupying the low-gap films of the junction bilayer, as pointed out in Refs.~\cite{sDiamond2022, sConnolly2024}. In Fig.~\ref{fig:Al_inj}(d) we diagram the junction film stack for $Q_1, Q_3, Q_5$. Here, the low-gap side of the junction is oriented towards 
the qubit island, and the high-gap side of the junction is connected to the ground plane. All qubits to the left of the feedline, as diagrammed in Fig.~\ref{fig:Al_inj}(a) (above the feedline in other device diagrams of the text), so, qubits $Q_1$, $Q_3$, and $Q_5$, have their junctions in this orientation. We will refer to this junction placement as the `non-ideal gap-engineered' placement. The qubits to the right of the feedline, as diagrammed in Fig.~\ref{fig:Al_inj}(c) (below the feedline in other device diagrams of the text), have their junctions deposited such that they are oriented with the low-gap side of the junction is oriented towards 
the ground plane and the high-gap side of the junction connected to the qubit island [Fig.~\ref{fig:Al_inj}(e)]. We will refer to this junction placement for qubits $Q_2$, $Q_4$, and $Q_6$ as the `gap-engineered' junction configuration. 

QPs near the gap edge within the $\sim185$-nm thick ($\Delta\sim$183~$\mu$eV) Al ground plane and the qubit island are suppressed from tunneling into either electrode of the junction due to the differences in the superconducting gap energies. Again, the junction consists of a bilayer with a bottom (top) film of thickness 40~nm (80~nm) with gap energy $\sim$195~$\mu$eV ($\sim$188~$\mu$eV). From Ref.~\cite{sConnolly2024}, we assume that the effective QP energy above the gap is given by $k_BT\sim2~\mu e$V 
for typical cryostat operating temperatures, and as a result, the QPs are suppressed from entering the junction electrodes. This superconducting gap energy difference contributes to the lower overall $\Gamma_p$ rates we observe for the Al ground plane device compared to the Nb ground plane device in Fig.~\ref{fig:phonon_only}(b) and Fig.~\ref{fig:PSD_fits}, in addition to the phonon downconversion via pair-breaking scattering within the ground plane of the device.

However, QPs may initially have some athermal energy distribution relative to the gap of the material given by $\delta E.$ From Ref.~\cite{sYelton2024}, QPs of energy $1.28\Delta$ in Al can diffuse as far as 200~$\mu$m before scattering to the gap edge, which is larger than the longest length scale of the qubit island geometry,$~\sim$100~$\mu$m. For athermal QPs of energy 1.04$\Delta<\Delta+\delta E\lesssim1.3\Delta$ within the thick Al qubit island/ground plane material, there will be a larger effective quasiparticle density within the qubit island due to its confined geometry relative to the larger volume of the ground plane. Additionally, these QPs have sufficient energy to tunnel into both bilayers of the qubit junction electrodes. 
For the non-ideal gap-engineered junction placement, QPs sourced from the qubit island may preferentially tunnel into the low-gap 80-nm thick film of the junction bilayer, which connects directly to the junction tunnel barrier resulting in enhanced $\Gamma_p$ and qubit relaxation. By contrast, for the gap-engineered junction placement, QPs sourced from the qubit island may also tunnel into the thicker 80-nm film of the junction bilayer. However, this film is not directly connected to the insulating tunnel barrier and is therefore suppressed from tunneling through the junction by the higher gap of the thinner 40-nm film by an Arrhenius factor $\sim e^{-\delta\Delta/\delta E}$, where $\delta\Delta$ is the gap difference between the 40-nm and 80-nm films of the junction: $\delta \Delta/h \approx$ 1.8~GHz. This asymmetry due to junction orientation with respect to the ground plane and the qubit island was also observed in Ref.~\cite{sHarrington2024}.
As a result, we expect a larger contribution of QP tunneling events for the non-ideal gap-engineered junction, where QPs in the qubit island diffusively transport into the low-gap side of the junction. This is what we observe in the charge-parity switching rates in Fig.~\ref{fig:PSD_fits}. Note that we would not expect this tunneling asymmetry to occur for differential qubits with a floating-style shunt capacitor, since the geometry on either side of the junctions would be nominally identical.
        
As we pointed out in Ref.~\cite{sYelton2024}, the characteristic diffusion length scale of Al is much larger than that of a Nb film, and as a result, we do not see this same junction placement dependence in the Nb ground plane film. Thus, depending on the material, accounting for the \textit{transport} of QPs within the superconducting film can be an important consideration in addition to the energy scale of the superconducting gap $\Delta$ in determining the 
QP poisoning dynamics.
        
We further test the presence of this effect by controllably poisoning the qubits by biasing an on-chip tunnel junction beyond $2\Delta_{\rm Al}/e$, where we emit pair-breaking phonons into the device substrate, identical to the phonon injection measurement protocols used in Ref.~\cite{sIaia2022, sYelton2024}, which involved devices with Nb ground planes. Here we measure a nominally identical device to the one used for $\Gamma_p$ measurements in Fig.~\ref{fig:PSD_fits}, with an Al ground plane.
For a range of idle times $\tau_i$ of a $T_1$ measurement sequence, we first perform a standard $T_1$ measurement pulse sequence with no injection pulse: $X$ pulse, idle for $\tau_i$, measure. Next, we use a similar pulse sequence, however, we bias an on-chip tunnel junction for 1~ms, then wait for some delay then apply an $X$ pulse, idle for $\tau_i$, then measure. These two pulse sequences allow us to fit a baseline $T^b_1$ with no phonon-mediated poisoning and a poisoned $T_1$. We use these quantities to determine the change in the qubit relaxation rate versus injection pulse delay $\Delta \Gamma_1 = 1/T_1-1/T_b.$ We can relate this measured quantity to a normalized quasiparticle density $x_{qp}$ in the junction: $x_{qp} = \pi \Delta \Gamma_1/\sqrt{2\Delta_{\textrm{Al}}\omega_{01}/\hbar}$, where $\omega_{01}$ is the qubit 0-1 transition frequency~\cite{sWang2014}. In Fig.~\ref{fig:Al_inj}(f-i) we show the results of this experiment for different delays. Note that we are also reporting negative delay values, which coincide with the $X$ pulse of the $T_1$ measurement occurring during the injection pulse, as done in Ref.~\cite{sYelton2024}. For these negative delay times, we end the injection pulse at the start of the $X$ pulse to avoid a further poisoning of the qubits during the subsequent $T_1$ idle times. We inject from two injector junction locations, one in the bottom left corner of the device, as diagrammed in Fig.~\ref{fig:Al_inj}(f-g), close to a `non-ideal gap-engineered' qubit $Q_1$, and one at the top right corner of the device, as diagrammed in Fig.~\ref{fig:Al_inj}(h-i), close to a `gap-engineered' qubit $Q_6$. The black lines in Fig.~\ref{fig:Al_inj}(f-i) coincide with a simplified version of the modeling presented in \cite{sYelton2024}, where we solve
\begin{equation}
    \frac{\textrm{d}x_{qp}}{\textrm{d}t} =-r x_{qp}^{2}-s x_{qp}+g(t),
\label{eq:xqp_model}
\end{equation}
where the QP recombination rate $r$ is set to $1/(10~{\rm ns})$, but, as mentioned in \cite{sYelton2024}, this term doesn't significantly contribute to the QP dynamics for the regime of $x_{qp}$ in these measurements. $s$ is the QP trapping rate due to variations in the superconducting gap~\cite{sWang2014}, 
potentially including the different gaps from the bilayer of the junction electrodes~\cite{sMarchegiani2022,sDiamond2022}. $g$ is the QP generation rate in the qubit junction electrodes. We simplify the modeling of Ref.~\cite{sYelton2024} by assuming that the generation rate at each qubit junction is a square pulse of 1~ms. We fit the amplitude of the generation rate and the trapping rate $s$, which are labeled in Fig.~\ref{fig:Al_inj}(f-i). We find for both injection locations the non-ideal gap-engineered junctions have a slower average trapping rate of $\sim0.18\times10^{-2}~\mu$s$^{-1}$ compared to the average trapping rate of $\sim0.67\times10^{-2}~\mu$s$^{-1}$ for the gap-engineered junctions. 
The difference in these trapping rates can be attributed to the junction orientation, following the same argument presented previously. The phonons emitted from the injector junction originate from QP recombination within both the 40-nm and 80-nm films of the injector junction bilayer. Therefore, the initial QP energies within the qubit island relative to the gap $\delta E$ will not be suppressed from entering the qubit junction electrodes, since their energy will be comparable to the gap energy of both junction electrode films on average. Additionally, the QP scattering rate is reduced for these QPs with energies just above the gap of the ground plane and qubit island. The average QP energy generated within the thick ground plane/qubit island material ($\Delta\sim$183~$\mu$eV) from the pair-breaking phonons generated from QP recombination within the 80-nm (40-nm) thick films of the injector junction bilayer, will be $188~\mu e$V ($195~\mu e$V). A QP of energy $188~\mu e$V ($195~\mu e$V) within a material of gap $\Delta\sim$183~$\mu$eV will have a characteristic diffusion length scale of $\sim$4~mm ($\sim$1~mm). These length scales are much longer than the largest dimensions of the qubit island of $\sim100~\mu$m but comparable to the larger size of the ground plane
~\cite{sYelton2024}.

\section { Details on charge-tomography measurements}
\label{sec:charge_tomo}

\begin{figure}[h!]
\centering
\includegraphics[width=\linewidth]{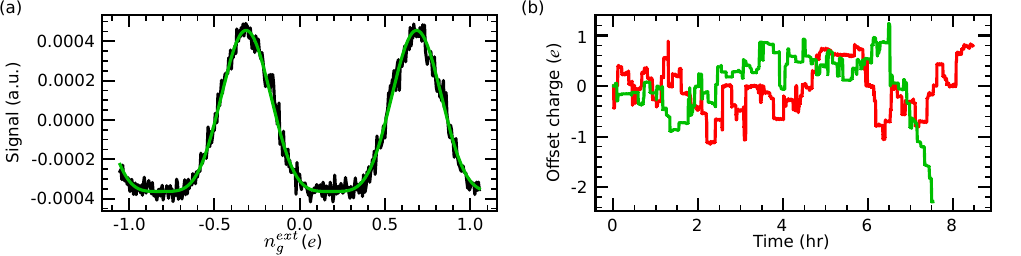}
\caption{
{\bf Charge tomography analysis details.} 
(a) Example charge tomography measurement for $Q_5$ of the Al ground plane sample in the conventional sample packaging. 
The averaged data are shown in the black and the resulting fit is shown in green. (b) Accumulated offset charge in time for the Al conventional sample (green) and the Nb conventional sample (red). For the duration of this experiment, the rate of discrete charge shifts $>0.15e$ defines $\Gamma_c$. 
\label{fig:CT_details}}
\end{figure}

As described in the main text, we can apply a dc voltage bias to a gate electrode near the qubit island, which allows us to vary the offset charge coupled to the qubit. For each bias point, we apply a similar pulse sequence as the QP charge-parity mapping protocol, described previously, but now we apply an $X/2$ pulse, idle for $1/2\delta f$, then apply another $X/2$ pulse. The final state along the qubit readout signal axis depends on the total offset charge coupled to the qubit, which is determined by both the nearby gate electrode as well as the environmental charge configuration. If the total offset charge coincides with the degeneracy point, no precession occurs during the idle phase and the qubit ends up in the 1 state. If the qubit is instead biased at the point of maximal charge dispersion, the state will precess by $\pi$, and the qubit ends in the 0 state after the final $X/2$ pulse. This protocol is independent of the QP charge-parity state. The resulting mapping to the qubit one-state occupation probability is given by: $P_1= \frac{1}{2} \left\{[d+\nu\cos[\pi\cos(2\pi n_g)]\right\}$, where $d$ and $\nu$ are fit parameters to the qubit signal axis and the total charge offset $n_g = n^{ext}_g+\delta n_{g}$, where $n^{ext}_g$ is the applied gate electrode charge and $\delta n_g$ is the environmental offset charge~\cite{sChristensen2019}. An example of this data (black) and a fit to $P_1$ is shown in Fig.~\ref{fig:CT_details}(a) for the Al ground plane sample in the conventional device packaging. 
        
We average multiple rounds of the dc bias sweep, typically $\sim 200$ rounds. We fit the resulting $P_1$ versus $n^{ext}_g$ 
for the averaged data and extract an environmental offset charge $\delta n_g$. We then repeat this for several hours and trace the accumulated offset charge versus time, as done in Ref.~\cite{sChristensen2019, sWilen2021}. Examples of the Nb ground plane and Al ground plane devices in the conventional sample packaging are shown in Fig.~\ref{fig:CT_details}(b). The rate of large charge jumps ($>$0.15e) of these data for multiple experiments throughout the cooldown is shown in Fig.~\ref{fig:phonon_only}(c) of the main manuscript.

\section{Correlated QP charge-parity switching}
\label{sec:corr_parity}

For the multi-qubit charge-parity measurements, we reject data where the environmental offset charge $\delta n_{g}$ was near the charge degeneracy point, where there is no precession about the equator and, thus, no determinable signal in the QP charge-parity mapping. We use the same 
offset-charge masking protocol described in Ref.~\cite{sLarson2025}. Note that we have selected a minimum state detection error of 15.7\% in order to build statistics for each point of the cooldown (details of state detection error 
discussed in Ref.~\cite{sLarson2025}). 
Part of the masking protocol involves applying a moving average to the raw data on a window of size $N_{w}$ to filter additive Gaussian white noise from the imperfect fidelity of the mapping sequence and the qubit readout.
We then fit a digital signal to the accepted and filtered data using a hidden Markov model (HMM), as done in Ref.~\cite{sIaia2022}. We next sweep a window that is $N_{w}$ data points long across the digital signals of the respective qubits; if there are multiple rising/falling edges on that window, we count this as a simultaneous QP charge-parity switch between the qubits involved~\cite{sIaia2022}. Here we pick $N_{w}$ so that during the duration of the cooldown the coincident window is a constant time interval $\Delta t_w.$ We set this time interval to be 1/4 of the shortest parity lifetime, $1/\Gamma_p$, observed for any qubit on a device for a given cooldown. 

\begin{figure}[h]
\centering
\includegraphics[width=\linewidth]{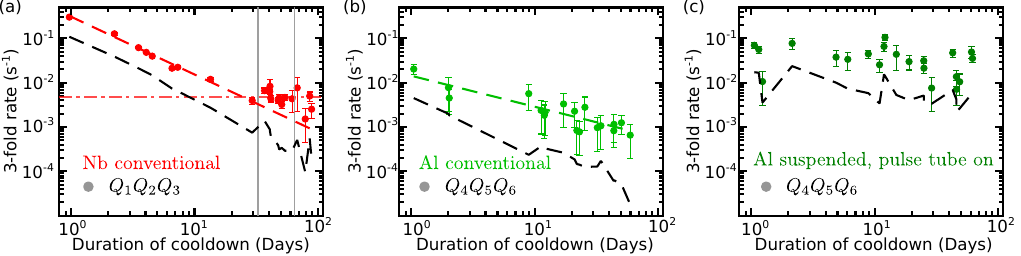}
\caption{
{\bf Three-fold coincident rates for the different device configurations.}
(a) Three-fold coincident parity measurements for qubits $Q1,Q2,Q3$ during the cooldown for the Nb ground plane device in the conventional sample package. 
Power law fit to the data is shown as a red dashed line. The ionizing radiation-limited saturation level at long  times is indicated as a dot-dashed horizontal red line. As in the main text, points where the cryostat rose above 100~mK are shown as the the transparent grey regions. (b) Three-fold coincident parity measurements for qubits $Q_4,Q_5,Q_6$ during the cooldown for the Al ground plane device in the conventional sample package. Power-law fit to the data is shown as a green dashed line. (c) Three-fold coincident parity measurements for qubits $Q_4,Q_5,Q_6$ during the cooldown for the Al ground plane device in the suspended sample package. 
Note that these measurements were taken while operating the cryostat with the pulse tube on. 
}
\label{fig:three_fold_compare}
\end{figure}

For an $n$-fold correlated event, the probability $p^{\rm random}_n$ for a random, uncorrelated $n$-fold coincidence occurring during the time interval $\Delta t_w$ is defined as the product of all $n$ single-fold switching probabilities for each qubit in the interval $t_w$. This single-fold switching probability for qubit $i$ is defined by $p_i = r_i \Delta t_w$, where $r_i$ is the parity switching rate 
for qubit $i$ from HMM analysis. As an example, the 3-fold random, uncorrelated probability for coincident switching of qubits $i,j,k$ is $p^{\rm random}_3 = r_i r_jr_k \Delta t_w^3$. We can also define a random, uncorrelated coincident switching rate for the interval $\Delta t_w$ by taking $r^{\rm random}_n = p^{\rm random}_n /\Delta t_w$. Our convention in the main text and in Fig.~\ref{fig:three_fold_compare} has been to show this random, uncorrelated coincident background rate as a dashed black line. 
        
In Fig.~\ref{fig:three_fold_compare} we compare the rates for three-fold coincident QP charge-parity switching for a similarly spaced group of qubits on the array for the three devices described in the main text; the Nb ground plane sample in the conventional sample packaging [Fig.~\ref{fig:three_fold_compare}(a)], the Al ground plane sample in the conventional sample packaging [Fig.~\ref{fig:three_fold_compare}(b)], and the Al ground plane sample in the suspended sample packaging [Fig.~\ref{fig:three_fold_compare}(c)]. Note that the data for the Al ground plane device in the suspended mount is taken while operating the pulse tube. The 3-fold coincident switching rate for the Nb ground plane sample decreases in time according to a power law of exponent $\alpha =-1.3(1)$ until it saturates to a rate of 4.8(5)$\times10^{-3}$s$^{-1}$. As described in the main text, we would expect the saturation level for an $n$-fold coincident switching rate for the Nb ground plane device to be approximately equal to $R_{\rm impact}/2^n$, where $R_{\rm impact}$ is the estimated rate of ionizing impacts for the entire device given the average charge jump rate $\Gamma_c$ [see Fig.~\ref{fig:phonon_only}(c)] and the charge-sensing area of our qubit islands from Ref.~\cite{sLarson2025}. We calculate that $R_{\rm impact}/2^3 = 4(1)\times10^{-3}$s$^{-1}$, approximately the saturation level of the observed 3-fold coincident event rate. Therefore, the Nb 3-fold events exhibit a similar saturation from ionizing impacts, consistent with the plot of 4-fold coincident switching rates presented in the main paper. We follow a similar analysis strategy for the Al conventional sample, shown in Fig.~\ref{fig:three_fold_compare}(b). Here we observe that the coincident 3-fold rate decreases in time according to a power law of exponent $\alpha = -0.7(2)$. We note that the exponent for this decrease is lower than we observe for the Nb ground plane device; also, there is no clear saturation at the longest times of this measurement. It is plausible that the gap-engineering of the ground plane thickness for the Al device has reduced the footprint of correlated events, causing a slower power-law decrease of the coincident switching rate in time. We leave the detailed characterization of the decrease of coincident switching rates for devices of different levels of gap-engineering as a topic for further study. Additionally, we do not see any clear power-law reduction over the cooldown duration for the 3-fold coincident QP charge-parity switching events for the Al ground plane device in the suspended packaging with the pulse tube on. Notably, the observed 3-fold rates with the pulse tube on are higher than for measurements of devices in the conventional sample packaging. For the pulse tube off measurements of the suspended device, we are unable to collect a sufficient amount of data in the 7-minute interval of the experiment to compute $n$-fold correlated parity switching rates.
        
\section{Sample box details}
\label{sec:sample_box}

\begin{figure}[h]
\centering
\includegraphics[width=\linewidth]{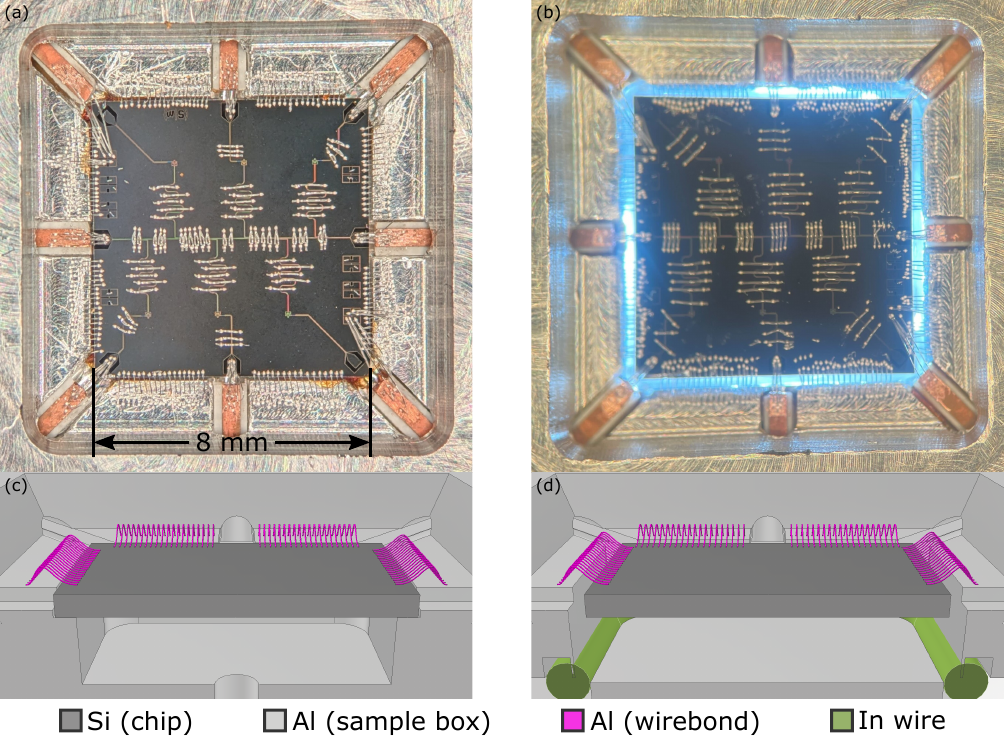}
\caption{
{\bf Conventional and suspended packaging details.}
(a) Stereoscope image of a nominally identical chip to the Al ground plane sample in the conventional package. (b) Stereoscope image of the Al ground plane sample in the suspended package with backlighting to highlight the excess gap between the chip and sample box. (c) and (d) are perspective cross-section views of the CAD models of the conventional and suspended packages, respectively, showing the chip (dark gray), sample box (light gray), and wirebonds (magenta). For the suspended package, indium wire (green) is used to seal the backing plate to the rest of the sample box once the wirebonding process is completed.
\label{fig:packaging_details}}
\end{figure}

Our sample boxes are machined out of Al with a pocket to house our 8~mm x 8~mm chips. The signal lines on the device are connected via wirebonds to Cu PCB traces for control, readout, and charge-biasing. We also connect to two test junctions in the top-left and bottom-right corners of the device for phonon injection experiments described in Sec.~\ref{sec:gap_engineering_Al} and Ref.~\cite{sIaia2022, sYelton2024}. In addition, roughly 40 wirebonds on each side of the chip also connect the device ground plane to the sample box for grounding. For the conventional sample package, small amounts of GE varnish anchor the chip to the ledge of the sample box. Figure~\ref{fig:packaging_details}(a) is a stereoscope image of the conventional packaging, and GE varnish (dark amber) can be seen at the corners of the sample. The ledge where the sample rests and is anchored with GE varnish can be seen in Fig.~\ref{fig:packaging_details}(c). The suspended packaging has this ledge machined away, and the pocket has been widened to allow the sample to be fully suspended by the ground wirebonds, even at millikelvin temperatures after accounting for the thermal contraction of the Al and Si [see Fig.~\ref{fig:packaging_details}(b) and (d)]. To wirebond the suspended sample, we use a temporary pillar that the supports the chip and sets the appropriate height of the device within the sample box. After wirebonding, this pillar is removed and a backing plate with an In-wire seal is used to maintain a light-tight enclosure, as in the conventional packaging.

    
\section{Pulse tube off measurements}
\label{sec:pulse_tube_off}

\begin{figure}[h]
\centering
\includegraphics[width=\linewidth]{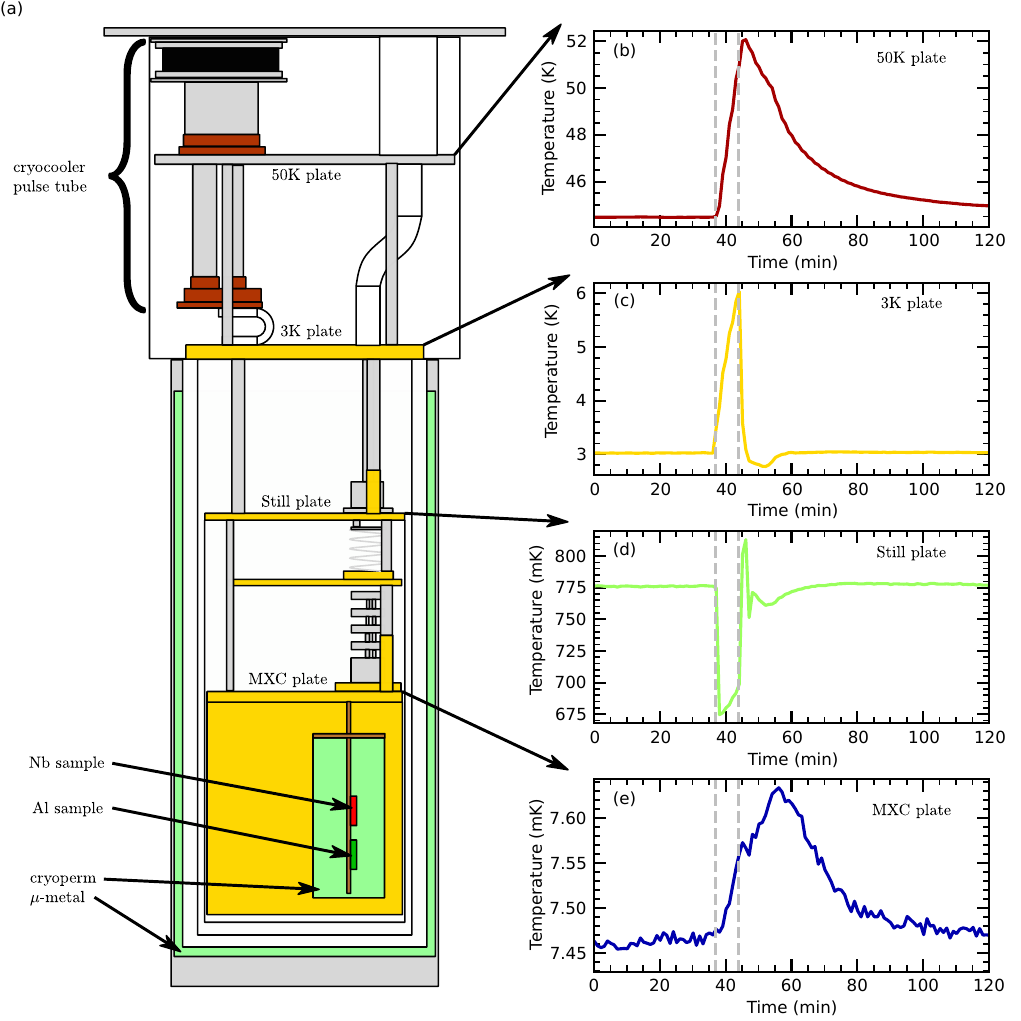}
\caption{
{\bf Cryostat response to turning off pulse tube.}
(a) Detailed diagram of the cryostat used in these experiments. The cryocooler pulse tube is off-axis of the center-line of vacuum jackets and heat shields. 
The samples are loaded at the mixing chamber (MXC) plate similarly, off-axis of the center-line of the cryostat. Note that the thermal-insulating supports and dilution unit are drawn. These components may create a coupling between the mechanical impulses of the cryostat and the qubit chip. The thermal response of the various stages of the cryostat when shutting off the pulse tube is shown for: (b) 50~K plate, (c) 3~K plate, (d) still plate, and (e) MXC plate. QP charge-parity switching data with the pulse tube on is taken at time=0. Next, the pulse tube is shut off (first vertical gray dashed line) and QP charge-parity data is acquired for $\sim$7 minutes. Finally, the pulse tube is turned back on (second vertical gray dashed line).
\label{fig:pulse_tube_temp}}
\end{figure}

As discussed in the main paper, we observe that the characteristic QP charge-parity switching rates $\Gamma_p$ for the device in the suspended sample packaging are sensitive to the pulse tube of the cryostat being on or off. We note that the study in Ref.~\cite{sKono2024} reported spurious qubit excitations that were synced with the cryocooler operation. In this case, GE varnish was used to mount the device, so the reported sensitivity to the pulse tube operation may be mediated by a different mechanical mechanism than what we observe in this work. Additionally, we did not observe spurious qubit excitations synced with the pulse tube operation for either the suspended or conventional packaged devices. In Fig.~\ref{fig:pulse_tube_temp}(a) we diagram the approximate configuration of the cryostat used throughout this work. The pulse tube, upper left of Fig.~\ref{fig:pulse_tube_temp}(a), is off-axis from the center-line of symmetry of the cryostat heat shields and vacuum jackets. The frequency of the cryocooler pulses is $\sim$1.4~Hz. The impulses from this cryostat component likely drive mechanical oscillations that propagate throughout the cryostat via the thermal-insulating support structures between the various stages of the cryostat [see Fig.~\ref{fig:pulse_tube_temp}(a)]. In the following section, we discuss a possible mechanical oscillation between the wirebonds and the suspended device that may act like a spring-mass system, a possible explanation for the sensitivity of QP charge-parity switching to pulse-tube operation that we observe. 

\begin{figure}[t!]
\centering
\includegraphics[width=\linewidth]{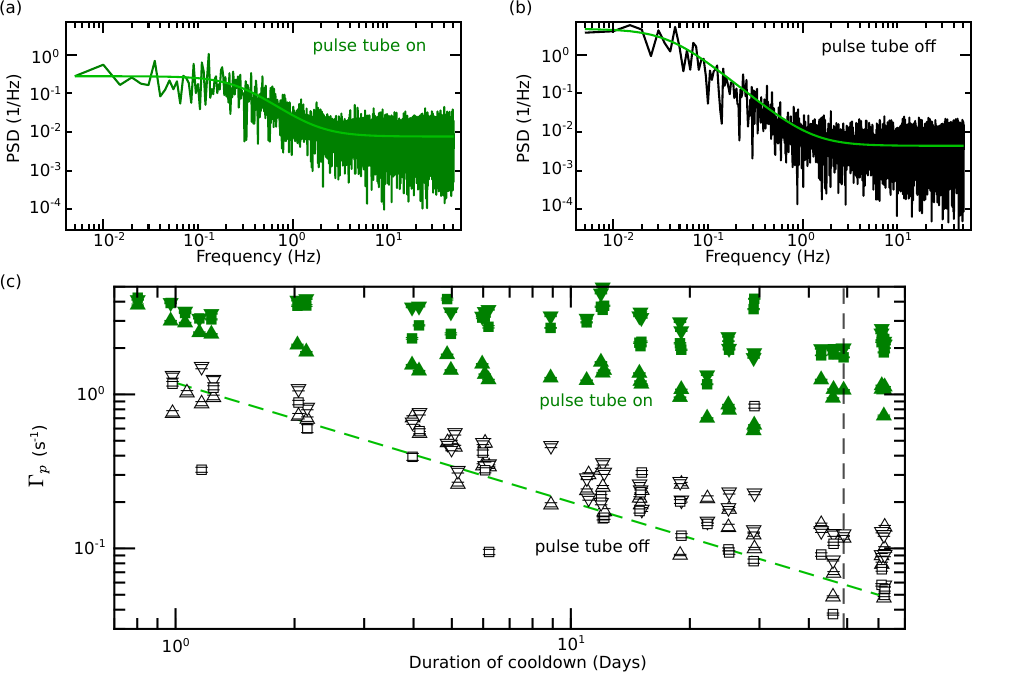}
\caption{
{\bf Pulse tube on/off experiments details during cooldown.} Example Lorentzian fits to QP charge-parity switching data of $Q_3$ for the Al suspended sample with the pulse tube on (a) and off (b). These measurements were both taken approximately 49 days into the cooldown. This point in time is shown as a dashed black vertical line in (c), which shows the characteristic QP charge-parity switching rates $\Gamma_p$ for three qubits $(Q_2, Q_3, Q_6)$ for the pulse tube on (dark green markers) and off (open black markers) experiments throughout the device cooldown. A power-law fit to the reduction of $\Gamma_p$ during the cooldown for $Q_6$ of the Al ground plane sample in the conventional sample package is plotted as a dashed green line for reference.
\label{fig:pulsetube_on_off_supp}}
\end{figure}

In performing the pulse tube on/off experiment, we found that the cryostat response limited the time over which we could keep the pulse tube off. However, we found that the cryostat response was repeatable using the following procedure. We first take a baseline measurement of $\Gamma_p$ with the pulse tube on 
at a repetition period of 1~ms for 50 rounds of 20,000 single-shots. This measurement sequence takes $\sim$30 minutes. Note that the beginning of this measurement is time $=0$ for the x-axis for the plots in Fig.~\ref{fig:pulse_tube_temp}(b-e). Immediately after this pulse sequence, we take another QP charge-parity measurement with a slower repetition period of 10~ms for 2 rounds of 20,000 single shots. This measurement sequence takes $\sim$7 minutes. The parameters of this experiment, including the repetition period, the number of single shots, and the number of rounds, are identical to the measurement sequence with the pulse tube off. Once this set of measurements is complete, we turn off the compressor to the pulse tube and turn off the cryostat heaters that regulate the still temperature. This is diagrammed as the first vertical dashed line in Fig.~\ref{fig:pulse_tube_temp}(b-e). For the 50~K, 3~K, and mixing chamber (MXC) plates, we see an increase in temperature at this point in the experiment, although this is quite modest for the MXC. However, the still plate drops in temperature due to the absence of applied heat. As soon as the pulse tube and still heaters are turned off, we initiate a QP charge-parity measurement sequence with 2 rounds of 20,000 single shots and a repetition period of 10~ms. The duration of this measurement is $\sim$7 minutes and is limited by the rise in pressure of the $^3$He circuit. If the experiment with the pulse tube off is run for much longer, $^3$He will begin to be returned to the external container for the $^3$He/$^4$He mixture, resulting in a significant rise in the cryostat temperature. Such an event was the source of the second temperature spike above 100~mK that is diagrammed in Fig.~\ref{fig:corr_parity}(a,b). As soon as this `pulse tube off' QP charge-parity measurement is complete, we turn the pulse tube back on and re-apply the heat to the still. This is noted as the second gray vertical dashed line in Fig.~\ref{fig:pulse_tube_temp}(b-e). For the 50~K and 3~K plates, we observe the temperatures begin to decrease back to their respective baseline operating levels [Fig.~\ref{fig:pulse_tube_temp}(b,c)]. There is a spike in the still temperature [Fig.~\ref{fig:pulse_tube_temp}(d)] due to the heat being reapplied to the plate. Soon after, the still recovers to its operating temperature synced with the recovery of the 3~K plate [Fig.~\ref{fig:pulse_tube_temp}(c,d)]. The MXC plate does not immediately begin to recover with turning the pulse tube back on, however, its maximum $\sim 3\%$ rise in temperature is negligible [Fig.~\ref{fig:pulse_tube_temp}(e)]. The 50~K plate limits the timing of the experiment since it takes the longest to recover to its baseline operating temperature at $\sim$1~hr [Fig.~\ref{fig:pulse_tube_temp}(b)].

We repeat this experimental procedure throughout the cooldown and summarize the resulting data in Fig.~\ref{fig:pulsetube_on_off_supp}. In Fig.~\ref{fig:pulsetube_on_off_supp}(a,b) we show example Lorentzian fits to the PSDs of the QP charge-parity switching data acquired with the pulse tube on/off procedure described above. These measurements are taken $\sim$49 days into the cooldown for $Q_3$; the dashed vertical line in Fig.~\ref{fig:pulsetube_on_off_supp}(c) coincides with this point in the cooldown. For the pulse tube on [off] we found the $\Gamma_p$ to be 1.07(3) [0.108(2)] s$^{-1}$. This order-of-magnitude difference between the pulse tube on or off is also observed for the other qubits on the device [Fig.~\ref{fig:pulsetube_on_off_supp}(c)]. We conducted a similar experiment for the Nb ground plane sample in the convention sample package at $\sim$49 days into the cooldown and found the $\Gamma_p$ for the pulse tube on [off] to be 0.476(1) [0.436(3)] s$^{-1}$ for $Q_3.$ For all qubits in this device there is no clear dependence on the pulse tube operation. However, for the sample in the suspended sample packaging, the pulse tube off QP charge-parity switching rates are lower, and these rates reduce in time following an inverse time power-law, similar to the same device in the conventional sample packaging. In Fig.~\ref{fig:pulsetube_on_off_supp}(c) we plot the fitted power-law dependence for $Q_6$ of the Al ground plane sample in the conventional configuration as a comparison. This agreement in the power-law reduction of the QP charge-parity switching rates regardless of the sample configuration points to a mechanism aside from the GE varnish mounting as the source of these phonon-only events. Additionally, the elevated rates observed in the suspended sample during the pulse tube operation are further evidence that these poisoning events are dominated by a mechanical process. In the following section, we propose a possible physical picture for the source of these mechanically induced phonon-only events for the suspended sample configuration. 

\section{Mechanical modes of the suspended sample}
\label{sec:mechanical_modes}

\begin{figure}[h]
\centering
\includegraphics[width=\linewidth]{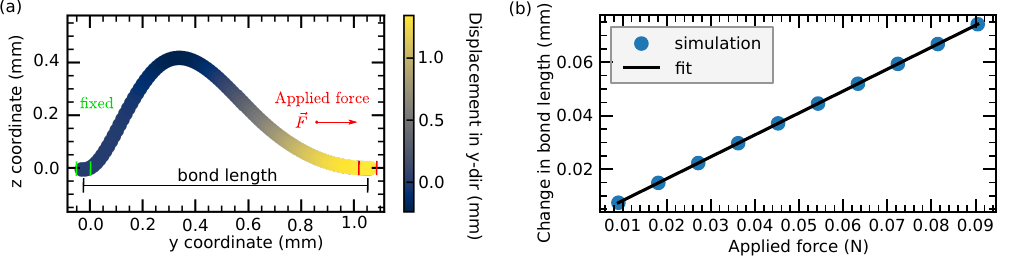}
\caption{
{\bf Simulation of wirebond effective spring constant.}
(a) Example linear elasticity simulation of a typical wirebond geometry using FEniCS. The left foot (green) is modeled to be fixed and a force of 0.09~N is applied in the $y$-direction to the second foot (red). The resulting displacement field in the y-direction along a line through the center of the wirebond geometry defines the color scale. (b) The change in distance between the two feet along the direction of the applied force is plotted for various force strengths. The scatter data represents different simulation results. The black line is a linear fit to the simulated data, giving an effective spring constant of $\approx1000$~N/m.
\label{fig:FEniCS_wirebond_sim}}
\end{figure}

It is plausible that the elevated charge-parity switching rates for the suspended sample configuration could be due to an effective spring-mass mechanical system that arises between the device and the perimeter of the wirebonds that anchor the device into the sample packaging. To quantify this effect, we simulate the effective spring constant of a single Al-1\%Si wirebond. To this end, we utilize an open-source finite element solver toolkit, FEniCS \cite{sAlnaes2015, sLogg2012}, to solve a variational form of linear elasticity equations for a typical wirebond geometry. We model the wirebond material as pure aluminum, using the elastic constants for Al at cryogenic temperatures from Ref.~\cite{sNamm1964}. The geometry can be seen in Fig.~\ref{fig:FEniCS_wirebond_sim}(a). The two `feet' of the wirebond are areas of the system where relevant simulation boundary conditions are enforced. The first foot [noted in green in Fig.~\ref{fig:FEniCS_wirebond_sim}(a)] coincides with the side of the wirebond that is bonded to the sample packaging and is assumed to be fixed in the simulation. The other foot [noted in red in Fig.~\ref{fig:FEniCS_wirebond_sim}(a)] is assumed to be attached to the device and is free to move. We apply a simulated force in the $y$-direction to the second foot, effectively stretching the bond, then calculate the displacement between the two feet of the wirebond along the direction of the applied force. We repeat this simulation for various force strengths; the results are plotted as blue scatter points in Fig.~\ref{fig:FEniCS_wirebond_sim}(b). This resulting linear relationship is the effective spring constant in the $y$-direction, $k_y \approx1000$~N/m. Doing a similar analysis for the other cartesian directions $z$, pushing the second foot up, $k_z \approx 200$~N/m, and $x$, pushing the second foot to the side [out of the page in Fig.~\ref{fig:FEniCS_wirebond_sim}(a)], $k_x \approx 100$~N/m. 

Given that the mass of the device is 0.08 grams 
, and that there are approximately 40 wirebonds on a side, the resonant frequency of the chip-wirebond side-to-side mode (displacement is orthogonal to the normal of the device) is $\sim6$~kHz. Note that this ignores bonds on the edge that are perpendicular to the direction of displacement. Similarly, for an in-and-out mode of the sample box pocket (displacement is along the normal of the device) the resonant frequency is $\sim3$~kHz. Thus, both modes resonate at a frequency of a few kHz, which is much higher than the frequency of the cryocooler pulse tube cycle of 1.4~Hz. Therefore, we do not assume that the pulse tube is driving the chip-wirebond system at a mechanical resonance. However, the suspension of the device via the wirebonds does create a source of elastic potential energy that is not present for the conventional sample packaging. Assuming the chip displaces by a small amount, say 1~nm, in a direction perpendicular to the normal of the device, the elastic energy stored in the wirebond-device spring-mass system is on the order of $10^{-14}$~J, or $10^{5}$~eV. Only a small fraction of that potential energy, on the order of $10^{-8}$, 
needs to dissipate at one of the wirebond contact points with the device to generate a pair-breaking phonon for Al. The details of this mechanical energy dissipation and possible dependence on the wirebond shape are subjects of further study. 


\input{bib_output_supp.tex}